# Thermodynamic properties of yttrium cuprate

N.I. Matskevich[1]*, Yu.F. Minenkov[2], G.A. Berezovskii[1]

[1]Nikolaev Institute of Inorganic Chemistry, Siberian Branch of the Russian Academy of Science, Prospect Acad. Lavrentiev, 3, 630090, Novosibirsk, Russian Federation

[2]INFN, Sezione di Roma Tor Vergata, I-00133 Roma, Italy

**Abstract**

The standard formation enthalpy and enthalpy from binary oxide of yttrium cuprate have been determined by solution calorimetry combining the solution enthalpies of $Y_2Cu_2O_5$ and $Y_2O_3 + 2CuO$ mixture in 6 M HCl at 323.15 K and literature data. The heat capacity of $Y_2Cu_2O_5$ has been measured by adiabatic calorimetry from 8 up to 303 K. Smoothed values of heat capacities, entropies and enthalpies were calculated on the basis of experimental data. The thermodynamic functions (heat capacity, entropy and enthalpy) at 298.15 K have been calculated as following: $C°_p$ (298.15 K) = 184.36 ± 0.07 J K$^{-1}$mol$^{-1}$; $S°$ (298.15 K) = 196.44 ± 0.56 J K$^{-1}$mol$^{-1}$; $H°$(298.15 K) – $H°$(0) = 31405 ± 40 J mol$^{-1}$. On the basis of obtained data it has been obtained that above-mentioned complex oxide is thermodynamically unstable with respect to their decomposition into binary oxides at room temperatures.

## 1. Introduction

High temperature superconductors (HTSC) which were discovered about 20 years ago are a subject of intensive investigations up to now [1-8]. It is very important to perform full physical and chemical study of systems containing HTSC compounds in order to understand the fundamental reasons of HTSC. Moreover, researchers and scientists have endeavored to improve the superconducting, mechanical, electrical, physical, thermodynamic, microstructural and flux pin-ning properties of the superconductor materials to make them suitable for high temperature and magnetic field applications. Among all their properties thermodynamic stability is of interest for many practical applications. The thermodynamic stability of complex oxides may have an impact on the mechanical stability of the microstructure of corresponding ceramics. There are a lot of reports on the determination of thermodynamic properties of compounds in HTSC systems. Here we would like to mention only several of them here [9-19].

---

*Corresponding author: Matskevich Nata I. nata@niic.nsc.ru, fax/phone: +7 383 3 30 64 49

In the present work we reported experimental data on formation enthalpy and heat capacity of yttrium cuprate ($Y_2Cu_2O_5$). This compound is related to HTSC. Solution calorimetry method was used to obtain formation enthalpy of $Y_2Cu_2O_5$. Low temperature adiabatic calorimetry was applied to measure heat capacity of yttrium cuprate in the temperature range of 8-303 K. Here we will notice that the same sample was used both for low temperature and solution calorimetry investigations. It is important because we will use experimental data to calculate free Gibbs energy.

## 2. Experimental part

*Preparation of sample*

Sample of $Y_2Cu_2O_5$ was prepared by solid-state synthesis heating stoichiometric amounts of dried $Y_2O_3$ (s) and CuO (s) (Cerac, mass fractions of compounds are more than 0.9999). Stoichiometric mixtures of $Y_2O_3$ and CuO were mixed and milled in a planetary mill with agate balls during 72 h with intermediate regrinding. Then the powders were pressed into pellets with a diameter of 10 mm and calcinated at 1270 K during 70 h. The compound was characterized by X–ray powder diffraction and chemical analysis. The content of all the metallic components was determined by atomic absorption [20, 21]. The oxygen content was determined by iodide titration using 0.01 N $Na_2S_2O_3 \cdot 5 H_2O$ according to the method described in paper [22]. According to the results of the analyses the involved compound was found to be single phase with an accuracy of about 1%.

*Experimental technique*

Solution calorimetry was chosen as a method to determine the formation enthalpy of $Y_2Cu_2O_5$. 6 M HCl was used as a solvent. Experiments were performed in an automatic solution calorimeter with isothermal shield. Construction of calorimeter and procedure of carrying out experiments were described earlier [23-25]. The main part of the calorimeter was a glass Dewar vessel (200 ml). The thermistor (MMT 4), calibration heater, mixer and the device to break the ampoules were mounted on the lid closing the Dewar vessel. A device, designed according to the CAMAC standard, was created to connect the calorimeter heater and thermistor with a computer. The reproducibility of the heat equivalent of the calorimeter was 0.03%. Dissolution of a standard substance (KCl) was performed to check the precision of the calorimeter. The obtained dissolution heat of KCl (17.529 ± 0.009 kJ/mol) is in a good agreement with the value recommended in the literature [26-27]. The experiments were performed at 323.15 K. At lower temperatures the rate of $Y_2O_3$ and $Y_2Cu_2O_5$ is small, which is not enough for precision measurements. The amounts of substances used were about 0.05-0.2 g. The identical state of the solution obtained by dissolution of $Y_2Cu_2O_5$ phase and dissolution of $Y_2O_3$ + 2 CuO mixture was proved by measuring the electronic spectra of these solution in the range of $10^4 – 3 \times 10^4$ cm$^{-1}$. It was shown that in experiments conducted in air, the spectra of the solutions became identical in 1-2 min after dissolution was finished.

Calorimetric cycle was designed in such a way that it was possible to determine the formation enthalpy of the 2:0:2 ($Y_2Cu_2O_5$) phase from yttrium oxide, copper oxide and then to calculate the standard formation enthalpy of yttrium cuprate on the basis of experimental and literature data.

The dissolution processes were described by the equation:

$$Y_2O_3 + \text{solution 1} = \text{solution 2} + \Delta_{sol}H^o_1 \qquad (1)$$

$$2CuO + \text{solution 2} = \text{solution 3} + 2\Delta_{sol}H^o_2 \qquad (2)$$

$$Y_2Cu_2O_5 + \text{solution 1} = \text{solution 3'} + \Delta_{sol}H^o_3 \qquad (3)$$

where solution 1 is 6 M HCl.

If solution 3 obtained after dissolution of $Y_2O_3$ + 2CuO mixture is identical to solution 3' obtained after dissolution of $Y_2Cu_2O_5$ is assumed to be identical the following equation can be written:

$$Y_2O_3 + 2CuO = Y_2Cu_2O_5 + \Delta_{ox}H^o_4 \qquad (4)$$

where $\Delta_{ox}H^o_4 = \Delta_{sol}H^o_1 + 2\Delta_{sol}H^o_2 - \Delta_{sol}H^o_3$.

Heat capacity of yttrium cuprate was measured using low temperature adiabatic calorimetry. The measurements were performed in a vacuum adiabatic calorimeter operated in periodic heat input mode. The calorimeter used here and the procedure of measurements have been described in detail elsewhere [28].

The sample of $Y_2Cu_2O_5$ was loaded into the calorimetric vessel. The calorimetric vessel (about 10 ml) was made of nickel. The vessel was hermetically sealed and hung on thin kapron wires into the space between two adiabatic shields. It was shown in paper [29] that two adiabatic shields were enough to achieve high precision adiabatic conditions. The accuracy of the temperature control of the heat shields was $10^{-3}$ K. The calorimeter was operated under vacuum ($10^{-4}$ atm).

The temperature of the calorimetric vessel was measured by a platinum resistance thermometer ($R_0 = 100.2608$ Ω). From 13.81 up to 305 K the temperatures were calculated using a standard function [30]. The sensitivity of the thermometer circuit was 2-5 x $10^{-5}$ K. The equation $R(T) = A + BT^2 + CT^5$ [31] was used to calculate temperature below 13.81 K. The calorimeter heater made of (Cu+Zn) alloy had a resistance of 300 Ω. The power of the heater was measured with a multimeter (DATRON 7061).

The heat capacity of the empty nickel vessel ($C_e$) was measured from 8 up to 305 K over 99 experimental points. The data $C_e = f(T)$ calculated on the basis of these experimental values were used to obtain the heat capacity of the investigated samples. The average deviation of the experimental heat capacities values from the smoothed $C_p$ curve for an empty vessel was 0.35% in the range 8-20 K and 0.03% in the range 20-305 K.

The accuracy of this calorimeter was tested by measuring the heat capacity of 6.0381 g of benzoic acid ($C_6H_5OOH$), recommended by the Calorimetry Conference (NBS-49) as a standard substance [32]. These measurements performed over 32 experimental points from 8 up to 305 K agree with literature data [33] to within 1% in the temperature range 8-30 and within less than 0.2% in the temperature range 30-305 K.

### 3. Results and discussion

The enthalpies of reactions (1)-(3) were: $\Delta_{sol}H^o_1$ (323.15 K) = -382.71 ± 1.84 kJ/mol [23], $\Delta_{sol}H^o_2$ (323.15 K) = -51.12 ± 2.13 kJ/mol [23], $\Delta_{sol}H^o_3$ (323.15 K) = -504.22 ± 3.77 kJ/mol. In this paper we measured only solution enthalpy of $Y_2Cu_2O_5$. Solution enthalpies of $Y_2O_3$ and CuO were taken from our earlier paper [23]. The measured enthalpies of dissolution were used for calculating the enthalpy of the reaction:

$$Y_2O_3 + 2CuO = Y_2Cu_2O_5 + \Delta_{ox}H^o_5 \qquad (4)$$

$\Delta_{ox}H^o_4$ (323.15 K) = 19.27 ± 5.16 kJ/mol

The dissolution enthalpies of $Y_2O_3$, CuO, $Y_2Cu_2O_5$ used for calculation of the enthalpy of reaction (4) were calculated as average values of six experiments. Errors were calculated for the 95% confidence interval using the Student coefficient.

Literature data for formation enthalpies of $Y_2O_3$ and CuO were used to calculate the formation enthalpy of $Y_2Cu_2O_5$ as following:

$\Delta_f H^o$(323.15 K) = -2209.4 ± 5.2 kJ/mol

Using heat capacities measured by us below we calculated the values for $Y_2Cu_2O_5$ at standard conditions:

$\Delta_{ox}H^o_4$ (298.15 K) = 19.35 ± 5.16 kJ/mol

$\Delta_f H^o$(298.15 K) = -2209.7 ± 5.2 kJ/mol

Comparison of data obtained by us on formation enthalpy from oxide for yttrium cuprate with literature data which was reviewed in Navrotsky's paper [9] showed that our data are in a good agreement with literature values.

To measure the heat capacity of $Y_2Cu_2O_5$, sample (weight, 7.45 g) was loaded into the nickel calorimetric vessel. It was necessary to crush the sample to a coarse powder to load into the calorimeter. 91 experimental values of $C_p$ were obtained. The obtained data are presented in Table 1. Smoothing of the experimental data of $Y_2Cu_2O_5$ heat capacities were treated using a computer programs described earlier [28]. Smoothed values of heat capacities, entropies and enthalpies are given in Table 2.

The average deviation of the experimental heat capacity values from the smoothed $C_p$ curve was about 0.5% in the temperature range 8-20 K, 0.1% in the temperature range 20-305 K.

The thermodynamic functions calculated under standard conditions are:

$C°_p$ (298.15 K) = 184.36 ± 0.07 J K$^{-1}$mol$^{-1}$;

$S°$ (298.15 K) = 196.44 ± 0.56 J K$^{-1}$mol$^{-1}$;

$H°(298.15 \text{ K}) - H°(0) = 31405 \pm 40 \text{ J mol}^{-1}$.

Reported uncertainties for $C_p$, $S$, $H$ were estimated by taking into account the average scatter of the experimental values of heat capacities.

Comparison of our data with data of paper [14] showed good agreement.

The experimental data of the heat capacities of $Y_2Cu_2O_5$ show that the sample has an anomalous rise at low temperatures (below 20 K). For the first time the anomaly of $C_p(T)$ was reported in paper [34]. We calculated the thermodynamic function of anomaly as following:

$\Delta_{tr}S = 5.1 \pm 0.7 \text{ J mol}^{-1} \text{ K}^{-1}$;

$\Delta_{tr}H = 51.2 \pm 3.7 \text{ J mol}^{-1}$.

Obtained data on formation enthalpy and heat capacity of yttrium cuprate allows us to understand if $Y_2Cu_2O_5$ will be stable at room temperature. On the basis of data of this paper we calculated the free Gibbs energy at 298.15 K for reaction (4) as following value:

$Y_2O_3 + 2CuO = Y_2Cu_2O_5 + \Delta_{ox}H°_5$ 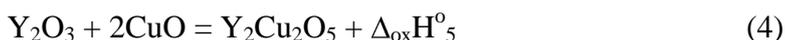 (4)

$\Delta_{ox}G°_4 = 15.83 \pm 5.17 \text{ kJ/mol}$

Calculation was performed using the program of Data Bank of Properties of Electronic Technique [35]. It means that yttrium cuprate is thermodynamically unstable at room temperature in respect to decomposition on mixture $Y_2O_3 + 2CuO$.

It is interesting to mention that entropy of reaction (4) is a small value:

$Y_2O_3 + 2CuO = Y_2Cu_2O_5 + \Delta_{ox}H°_4$ 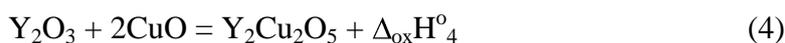 (4)

$\Delta_{ox}S°_4 = 11.8 \text{ J/K mol}$

It means that entropy of mixed oxide can be estimated as sum of entropies of binary oxides.

**Conclusion**

We measured the standard formation enthalpy of $Y_2Cu_2O_5$ by solution calorimetry in 6 M HCl. We measured heat capacity of yttrium cuprate by low temperature adiabatic calorimetry in the temperature range 8-303 K. Smoothed values of heat capacities, entropies and enthalpies were calculated on the basis of experimental data. We determined the stability of yttrium cuprate with respect to mixture of binary oxides. On the basis of our data we established that above-mentioned complex oxide is thermodynamically unstable with respect to their decomposition into binary oxides at room temperatures.

**Acknowledge**

This work is supported by Karlsruhe Institute of Technology (Germany), RFBR (project 13-08-00169) and Program of Fundamental Investigation of Siberian Branch of the Russian Academy of Sciences.

Table 1. Experimental values of heat capacities of $Y_2Cu_2O_5$ (molar mass 384.901)

| T/K | $C_p$/JK$^{-1}$mol$^{-1}$ | T/K | $C_p$/JK$^{-1}$mol$^{-1}$ | T/K | $C_p$/JK$^{-1}$mol$^{-1}$ | T/K | $C_p$/JK$^{-1}$mol$^{-1}$ |
|---|---|---|---|---|---|---|---|
| 8.67 | 4.589 | 90.06 | 67.74 | 187.42 | 142.4 | 212.67 | 155.1 |
| 9.31 | 5.148 | 91.38 | 68.84 | 190.88 | 144.2 | 214.11 | 155.7 |
| 10.08 | 5.969 | 95.79 | 73.21 | 191.11 | 144.3 | 214.66 | 155.9 |
| 11.00 | 5.097 | 101.21 | 78.19 | 194.32 | 145.9 | 215.94 | 156.8 |
| 12.27 | 3.735 | 106.40 | 82.87 | 194.86 | 146.3 | 218.22 | 158.0 |
| 14.33 | 3.497 | 111.39 | 87.35 | 197.39 | 147.1 | 219.61 | 158.0 |
| 17.52 | 4.125 | 114.99 | 90.31 | 197.44 | 147.2 | 221.75 | 159.1 |
| 19.88 | 4.911 | 116.22 | 91.41 | 197.55 | 147.3 | 225.71 | 160.7 |
| 22.47 | 6.085 | 120.72 | 95.18 | 199.50 | 148.5 | 226.14 | 161.3 |
| 25.88 | 8.273 | 126.27 | 99.62 | 200.00 | 148.9 | 231.75 | 163.3 |
| 29.58 | 10.89 | 131.67 | 104.2 | 200.98 | 149.4 | 237.75 | 165.7 |
| 33.63 | 14.04 | 136.94 | 108.2 | 202.44 | 150.2 | 243.70 | 167.9 |
| 38.13 | 18.42 | 142.51 | 112.5 | 202.45 | 150.1 | 249.61 | 169.9 |
| 43.15 | 22.81 | 148.61 | 116.9 | 203.92 | 150.9 | 255.48 | 172.3 |
| 49.12 | 28.34 | 154.83 | 121.7 | 204.87 | 151.6 | 261.31 | 174.1 |
| 55.10 | 34.10 | 160.93 | 125.8 | 205.39 | 151.8 | 267.10 | 176.1 |
| 61.06 | 39.93 | 166.91 | 129.7 | 206.85 | 152.7 | 272.85 | 177.8 |
| 67.74 | 46.41 | 172.80 | 133.4 | 207.68 | 153.0 | 278.57 | 179.8 |
| 74.13 | 52.72 | 178.08 | 136.9 | 208.31 | 153.3 | 284.52 | 181.4 |
| 79.88 | 58.20 | 178.60 | 137.2 | 209.77 | 153.8 | 290.71 | 183.1 |
| 82.28 | 60.00 | 180.92 | 138.6 | 209.85 | 153.7 | 296.87 | 184.1 |
| 85.18 | 63.25 | 184.06 | 140.5 | 211.08 | 154.4 | 302.99 | 185.1 |
| 86.96 | 64.54 | 184.71 | 140.8 | 211.22 | 154.5 | | |

Table 2. Smoothed values of the thermodynamic functions of $Y_2Cu_2O_5$ (molar mass 384.901)

| T/K | $C_p$/ JK$^{-1}$mol$^{-1}$ | S/ JK$^{-1}$mol$^{-1}$ | H(T)-H(0)/ Jmol$^{-1}$ | T/K | $C_p$/ JK$^{-1}$mol$^{-1}$ | S/ JK$^{-1}$mol$^{-1}$ | H(T)-H(0)/ Jmol$^{-1}$ |
|---|---|---|---|---|---|---|---|
| 5 | 0.7104 | 0.8433 | 1.924 | 160 | 125.17 | 98.855 | 9319.0 |
| 10 | 1.6332 | 1.5581 | 7.912 | 165 | 128.45 | 102.76 | 9953.1 |
| 15 | 2.9951 | 2.4994 | 19.83 | 170 | 131.66 | 106.64 | 10603 |
| 20 | 5.0863 | 3.6475 | 40.02 | 175 | 134.88 | 110.50 | 11270 |
| 25 | 7.8376 | 5.0640 | 72.01 | 180 | 138.03 | 114.35 | 11952 |
| 30 | 11.325 | 6.7907 | 119.6 | 185 | 140.99 | 118.17 | 12650 |
| 35 | 15.495 | 8.8424 | 186.4 | 190 | 143.72 | 121.97 | 13362 |
| 40 | 20.015 | 11.204 | 275.1 | 195 | 146.22 | 125.73 | 14086 |
| 45 | 24.558 | 13.824 | 386.5 | 200 | 148.83 | 129.47 | 14824 |
| 50 | 29.225 | 16.652 | 520.9 | 205 | 151.53 | 133.17 | 15575 |
| 55 | 34.027 | 19.661 | 679.0 | 210 | 153.94 | 136.86 | 16339 |
| 60 | 38.888 | 22.830 | 861.3 | 215 | 156.22 | 140.50 | 17114 |
| 65 | 43.755 | 26.135 | 1067.9 | 220 | 158.41 | 144.12 | 17901 |
| 70 | 48.644 | 29.556 | 1298.9 | 225 | 160.56 | 147.71 | 18698 |
| 75 | 53.514 | 33.079 | 1554.3 | 230 | 162.66 | 151.26 | 19506 |
| 80 | 58.311 | 36.686 | 1833.9 | 235 | 164.60 | 154.78 | 20325 |
| 85 | 62.957 | 40.361 | 2137.1 | 240 | 166.53 | 158.26 | 21152 |
| 90 | 67.605 | 44.090 | 2463.4 | 245 | 168.34 | 161.72 | 21990 |
| 95 | 72.376 | 47.873 | 2813.4 | 250 | 170.14 | 165.13 | 22836 |
| 100 | 77.059 | 51.705 | 3187.0 | 255 | 171.98 | 168.52 | 23691 |
| 105 | 81.601 | 55.575 | 3583.7 | 260 | 173.73 | 171.88 | 24556 |
| 110 | 86.045 | 59.475 | 4002.9 | 265 | 175.35 | 175.20 | 25428 |
| 115 | 90.357 | 63.395 | 4443.9 | 270 | 176.94 | 178.50 | 26309 |
| 120 | 94.531 | 67.329 | 4906.2 | 275 | 178.54 | 181.76 | 27198 |
| 125 | 98.637 | 71.271 | 5389.1 | 280 | 180.10 | 184.99 | 28094 |
| 130 | 102.75 | 75.220 | 5892.6 | 285 | 181.51 | 188.19 | 28998 |
| 135 | 106.75 | 79.173 | 6416.4 | 290 | 182.77 | 191.36 | 29909 |
| 140 | 110.57 | 83.125 | 6959.8 | 295 | 183.77 | 194.49 | 30826 |
| 145 | 114.29 | 87.070 | 7522.0 | 300 | 184.62 | 197.59 | 31747 |
| 150 | 117.99 | 91.007 | 8102.6 | 305 | 185.53 | 200.64 | 32672 |
| 155 | 121.67 | 94.936 | 8701.8 | | | | |